\documentclass[12pt]{article}
\usepackage{amsmath}
\usepackage{amsthm}
\usepackage{amssymb}
\usepackage[english]{babel}
\usepackage[dvips]{graphicx}

\newcommand {\bs}{\boldsymbol}

\begin{document}

\title{One class of solutions with two invariant relations\\
for the problem of motion of the Kowalevski top\\in double constant
field\thanks{First published in Russian: Mekh. Tverd. Tela 32(2002),
pp. 32-38}}

\author {Mikhail P.\,Kharlamov}
\date{05.11.2002}

\maketitle

\begin{abstract}
Consider a rigid body having a fixed point in a superposition of two
constant force fields (for example, gravitational and magnetic).
Introducing the condition of Kowalevski type, O.I.\,Bogoyavlensky
(1984) has found the first integral generalizing that of Kowalevski
and pointed out the integrable case with two invariant relations,
which reduces to the 1st Appelrot class when one of the fields
vanishes. The article presents a new case with two invariant
relations integrable in Jacobi sense and generalizing the 2nd and
3rd classes of Appelrot.
\end{abstract}

{\bf Introduction.} Investigating the problems of classical
mechanics with $n$ degrees of freedom, one distinguishes the notions
of Liouville and Jacobi integrability. In the first case we have $n$
independent first integrals in involution and the corresponding
Hamiltonian system of $2n$ ordinary differential equations can be
derived (in theory) to a "simple"\, flow on an $n$-dimensional
surface. The second case appears when the equations have $2n - 2$
independent first integrals. Then the solution of the problem is
reduced to integrating of two differential equations having Jacobi's
last multiplier (in contemporary terms---to the system with
invariant measure on a two-dimensional torus). Obviously if $n = 2$
these two cases are the same. The formal equivalence of the notions
takes place if among $n$ involutive integrals one has $n - 2$
integrals generated by symmetries of the potential and kinetic
energy. Then by ignoring the corresponding features of motion the
problem is reduced to a Hamiltonian system with two degrees of
freedom of natural structure (the phase space is the
coordinate--velocity space, the Lagrange function is quadratic
w.r.t. velocity components). The typical integral manifolds of the
reduced system are two-dimensional tori and the trajectories are
quasi-periodic.

If we really want to imagine the motion in all details, then, as a
result of our three-dimensional way of reasoning, we prefer the
Jacobi integrability. Indeed in the case of symmetries only the
solutions of the reduced system are double-periodic. The motions of
the initial system remain essentially $n$-dimensional. Let us for
brevity call this situation a reducible problem. Jacobi
integrability means that the trajectories of the initial mechanical
system fill some two-dimensional surfaces, which can be easily
represented in ${\bf{R}}^3$. The projections of such trajectories
onto the spaces having physical sense (say, the hodographs of the
angular velocity or the traces of the vertical vector in the moving
frame) can be studied in reality.

The general integrability cases in the dynamics of a rigid body
belong to the reducible problems. They are conditionally interpreted
as two-dimensional by ignoring the precession part of motion. But
the trajectories describing the evolution of the orientation matrix
in the general case still fill a three-dimensional torus. This is
why the complete classification of the immovable hodographs of the
angular velocity and based on it complete geometrical simulation of
motion seem to be extremely complicated problems. Moreover,
considering non-symmetric force fields we loose hope to find the
possibility of any natural reduction to two-dimensional
configuration spaces.

Thus, much attention is payed to the cases of Jacobi integrability
of the full system of equations of a rigid body motion. These are
the cases when the integral manifolds of this system are
two-dimensional. For the reducible problems the corresponding
solutions of the Euler--Poisson equations would be singular periodic
motions of elliptic or hyperbolic type. Such solutions are called
partial. Given the body characteristics, the partial solution may be
the set of closed orbits representing, in the reduced system, a
family of two-periodic motions of the body. Many partial solutions
were found lately with the method of invariant relations
\cite{bib1}. To describe a partial solution (one-dimensional
invariant manifold in 6-dimensional space of the Euler--Poisson
variables) one has to find, in addition to three known energy, area,
and geometrical integrals, two independent invariant relations.

If the force field has no symmetries, then the full system of
Euler--Poisson equations contain 12 variables and admit 7
independent integrals, namely, the energy integral and 6 geometrical
integrals. Therefore, to obtain the Jacobi integrability one has to
find three invariant relations. In this paper we present such a
solution for the problem of the rigid body motion in two constant
fields.

{\bf 1.\,Main equations.} Consider the problem of motion of a rigid
body about a fixed point in the potential force field with a force
function of the type
\begin{equation}\label{eq1}
({\bf e}_1 ,{\bs \alpha}) + ({\bf e}_2 ,{\bs \beta}).
\end{equation}
Here $(\cdot , \cdot)$ denote the scalar product, the vectors ${\bf
e}_1 ,{\bf e}_2 $ are fixed in the body, and ${\bs \alpha}, {\bs
\beta}$ are immovable in inertial space. If ${\bs \beta} = 0$ (or
${\bs \alpha} \times {\bs \beta} = 0$, which is the same) we come to
the classical problem of motion of a heavy rigid body. The force
function (\ref{eq1}) arises, for example, in the problems of motion
of a magnetized body with a fixed magnetic momentum in constant
gravitational and magnetic fields, or an electrically charged body
(with immovable charges in it) in constant gravitational and
electric fields. Given that ${\bs \alpha} \times {\bs \beta} \ne 0$,
the orientation matrix of the body and acting forces are completely
defined by the components, with respect to the moving frame, of the
pair ${\bs \alpha},{\bs \beta}$. Therefore, the configuration space
of the problem (the group of the orthogonal $3\times 3$-matrices)
cannot be reduced to a space of less dimension, and this is a
principal difference with the case of one field. The corresponding
equations
\begin{eqnarray}
& {\bf I} \dot {\bs \omega} = {\bf I}{\bs \omega } \times {\bs
\omega} + {\bf e}_1 \times {\bs \alpha} + {\bf e}_2 \times {\bs
\beta},\label{eq2}\\
& \dot {\bs \alpha}   = {\bs \alpha} \times {\bs \omega}, \quad \dot
{\bs \beta} = {\bs \beta} \times {\bs \omega} \label{eq3}
\end{eqnarray}
can be considered as equations in ${\bf R}^9$ with three geometrical
integrals
\begin{equation}\label{eq4}
({\bs \alpha},{\bs \alpha}) = a^2 , \quad ({\bs \beta},{\bs \beta })
= b^2 , \quad ({\bs \alpha},{\bs \beta}) = c.
\end{equation}
Eqs.~(\ref{eq2}),(\ref{eq3}) have the energy integral
$$
H = \frac{1}{2} ({\bf I}{\bs \omega},{\bs \omega}) - ({\bf e}_1,
{\bs \alpha}) - ({\bf e}_2, {\bs \beta}).
$$
In general case, there is no linear integral of the area integral
type.

Choose the principal axes of the inertia tensor ${\bf I}$ for the
moving frame, then
$$
{\bf I}= \mathop{\rm diag}\nolimits(A_1, A_2,A_3).
$$
In the sequel, it is convenient to consider the vectors ${\bf e}_1,
{\bf e}_2$ orthonormal, and include all characteristic multipliers
into the parameters  $a, b, c$ of relations (\ref{eq4}). Take $\sqrt
{A_3 /u_0 }$ as the time unit ($u_0$ is some common unit of
measurement for the components of ${\bs \alpha}, {\bs \beta}$).
Formally it is equivalent to the choice $A_3=1$. We can use $u_0$ to
bring one of the constants $a, b, c$ to 1. However, we prefer to
keep notation (\ref{eq4}) for some natural symmetry in the formulas
below.

In the analogue of the Kowalevski case
$$
A_1  = A_2  = 2A_3 , \quad {\bf e}_1  = (1,0,0), \quad {\bf e}_2  =
(0,1,0).
$$
the Euler equations take the form
\begin{equation}\label{eq5}
2\dot \omega _1 = \omega _2 \omega _3  + \beta _3, \quad 2 \dot
\omega _2 =  - \omega _1 \omega _3  - \alpha _3 , \quad \dot \omega
_3 = \alpha _2  - \beta _1 .
\end{equation}
They are closed by the Poisson equations (\ref{eq3}).

O.I.\,Bogoyavlensky \cite{bib2} showed that Eqs.~(\ref{eq3}),
(\ref{eq5}) have the first integral of the Kowalevski type
\begin{equation}\label{eq6}
K = J_1^2  + J_2^2 ,
\end{equation}
where
$$
J_1  = \omega _1^2  - \omega _2^2  + \alpha _1  - \beta _2, \quad
J_2  = 2\omega _1 \omega _2  + \alpha _2  + \beta _1,
$$
and pointed out that on the zero level of the integral~(\ref{eq6}),
\begin{equation}\label{eq7}
J_1  = 0,\quad J_2  = 0,
\end{equation}
there exists a new partial integral, namely,
$$
J_3  = (\omega _1^2  + \omega _2^2 )\omega _3 + 2(\omega _1 \alpha
_3  + \omega _2 \beta _3 ),
$$
the constant of which is arbitrary.

Thus, the system of invariant relations (\ref{eq7}) defines the
four-dimensional manifold $M^4$; this manifold is independent of
integration constants. The induced (not reduced) system on it has
two first integrals
\begin{equation}\label{eq8}
H = h,\quad J_3  = j
\end{equation}
with arbitrary constants $h, j$. Therefore the initial system has a
partial case of Jacobi integrability. The system thus obtained on
$M^4$ can be represented in the Hamiltonian form \cite{bib2}, but
$M^4$ does not have a structure of a phase space of mechanical
system (coordinate-velocity structure). The topology of $M^4$ and of
two-dimensional integral manifolds (\ref{eq8}) is studied in the
work \cite{bib3}. The Bogoyavlensky solution generalized the
classical case of N.B.\,Delone.

{\bf 2.\,New solution with two invariant relations.} Below the term
"derivative" will mean differentiating functions of the variables
$\omega _i ,\alpha _j ,\beta _k $ in virtue of Eqs.~(\ref{eq3}),
(\ref{eq5}), i.e., the time-derivative along trajectories.

Note that the derivative, in the above sense, of any expression not
containing $\omega_3$, is linear in $\omega_3$. Suppose that some
function $F$ is linear in $\omega_3$ and its derivative does not
depend on $\omega_3$. Then there is a chance that the second
derivative of $F$, being linear in $\omega_3$, can appear to be
proportional to $F$, thus closing the sequence of differentiations
and generating the invariant relation in the definition of the
work~\cite{bib1}.

Following the idea of S.\,Kowalevski \cite{bib4} of using complex
variables, introduce the change of variables ($i^2=1$):
\begin{equation}\label{eq9}
\begin{array}{ll}
x_1 = (\alpha_1  - \beta_2) + i(\alpha_2  + \beta_1),&
x_2 = (\alpha_1  - \beta_2) - i(\alpha_2  + \beta_1 ), \\
y_1 = (\alpha_1  + \beta_2) + i(\alpha_2  - \beta_1), & y_2 =
(\alpha_1  + \beta_2) -
i(\alpha_2  - \beta_1), \\
 z_1 = \alpha_3  + i\beta_3, &
z_2 = \alpha_3  - i\beta_3,\\
w_1 = \omega_1  + i\omega_2 , & w_2 = \omega_1  - i\omega_2.
\end{array}
\end{equation}
Denoting the derivative with respect to $\tau=it$ by the prime, we
obtain from (\ref{eq3}), (\ref{eq5})
\begin{equation}\label{eq10}
\begin{array}{c}
\begin{array}{ll}
{x'_1  =  - x_1 w_3  + z_1 w_1,} & {x'_2  = x_2 w_3  - z_2 w_2,} \cr
{y'_1  =  - y_1 w_3  + z_2 w_1,} & {y'_2  = y_2 w_3  - z_1 w_2 ,}
\cr {2z'_1  = x_1 w_2  - y_2 w_1,} & {2z'_2  =  - x_2 w_1 + y_1
w_2,}
\end{array} \\
2w'_1  =  - (w_1 w_3  + z_1 ),\quad 2w'_2  = w_2 w_3  + z_2, \quad
2\omega'_3 = y_2  - y_1.
\end{array}
\end{equation}

Let
\begin{equation}\label{eq11}
\theta  = x_1 x_2 ,\quad Q_1  = x_2 z_1 w_1  + x_1 z_2 w_2 , \quad
Q_2 = x_2 z_1 w_1  - x_1 z_2 w_2.
\end{equation}
Then $ \theta ' = Q_2$, $Q'_1  = \frac{1}{2}Q_2 \omega _3 + ...$,
where "$...$"\, stands for the terms not depending on $\omega_3$.
Construct the combination
$$
\theta ^m \omega _3  - \theta ^n Q_1.
$$
In its derivative, the coefficient of $\omega_3$ is equal to
$$
m\theta ^{m - 1} \theta ' - \frac{1}{2}\theta ^n Q_2  = (m\theta ^{m
- 1}  - \frac{1}{2} \theta ^n )Q_2
$$
and vanishes if $m = \frac{1}{2},n =  - \frac{1}{2}$. Therefore, the
following function becomes "suspicious"\, in the sense of invariant
relation generating,
$$
F_1  = \sqrt {x_1 x_2 } \omega _3  - \frac{1} {\sqrt {x_1 x_2 } }
(x_2 z_1 w_1  + x_1 z_2 w_2 ).
$$

Calculate its derivative in virtue of (\ref{eq10}),
\begin{equation}\label{eq12}
\frac{d F_1} {d\tau}  = \frac{1} {2\sqrt {x_1 x_2 }} \,[\frac{x_2}
{x_1 } (z_1^2  + x_1 y_2 )(w_1^2  + x_1 ) - \frac{x_1} {x_2 } (z_2^2
+ x_2 y_1 )(w_2^2  + x_2 )].
\end{equation}
Note that the geometrical integrals (\ref{eq4}) imply
$$
\begin{array}{l}
z_1^2  + x_1 y_2  = (a^2  - b^2 ) + 2ic = c_1  = {\rm const},\\
z_2^2  + x_2 y_1  = (a^2  - b^2 ) - 2ic = c_2  = {\rm const}.
\end{array}
$$
Introduce the following notation
\begin{equation}\label{eq13}
U_1  = \frac{x_2 } {x_1 } c_1 (w_1^2  + x_1 ),\qquad U_2  =
\frac{x_1 } {x_2 } c_2 (w_2^2  + x_2 ),\qquad U_2  = \overline {U_1
}
\end{equation}
and calculate the derivatives of (\ref{eq13}),
$$
\begin{array}{l}
\displaystyle{\frac{dU_1} {d\tau } =\phantom{-}{\frac {c_1 } {x_1^2 }}(w_1^2  + x_1 )[x_1 x_2 \omega _3  - (x_2 z_1 w_1  + x_1 z_2 w_2 )],}\\
\displaystyle{\frac{dU_2} {d\tau } =-{\frac {c_2 } {x_2^2 }}(w_2^2 +
x_2 )[x_1 x_2 \omega _3  - (x_2 z_1 w_1  + x_1 z_2 w_2 )],}
\end{array}
$$
whence
\begin{equation}\label{eq14}
\begin{array}{l}
\displaystyle{\frac {d} {d\tau } (U_1  - U_2 ) = \frac{1} {\sqrt
{x_1 x_2 } }[{\frac {c_1 } {x_1^2 }}(w_1^2  + x_1 ) + {\frac{c_2 }
{x_2^2 }}(w_2^2  + x_2 )] (\sqrt {x_1 x_2 } \omega _3  - {\frac{x_2
z_1 w_1  + x_1 z_2 w_2 } {\sqrt {x_1 x_2 } }}).}
\end{array}
\end{equation}
Denote $F_2  = U_1  - U_2$ and rewrite (\ref{eq12}) and (\ref{eq14})
in the form
$$
\begin{array}{l}
\displaystyle{\frac {dF_1} {d\tau }   = \frac {1} {2\sqrt {x_1 x_2 }
}\, F_2 , \quad \frac {dF_2} {d\tau }   = \frac {1} {\sqrt {x_1 x_2
} }(U_1  + U_2 )\,F_1.}
\end{array}
$$
Hence the system of relations
\begin{equation}\label{eq15}
F_1  = 0,\quad F_2  = 0
\end{equation}
defines the invariant submanifold of the phase space of
Eqs.~(\ref{eq10}). Given the geometrical identities~(\ref{eq4})
depending only on the body parameters and the force fields, this
manifold has dimension~4. Denote it by $N^4$.

Note that the expressions $\theta$ and $Q_1$ in (\ref{eq11}),
(\ref{eq13}) are real, while $U_1-U_2$ is purely imaginary. Hence
Eqs.~(\ref{eq15}) can be written in the form
$$
x_1 x_2 \omega _3  - 2{\mathop{\rm Re}\nolimits} (x_2 z_1 w_1 ) =
0,\qquad {\mathop{\rm Im}\nolimits} [x_2^2 c_1 (w_1^2  + x_1 )] = 0.
$$
Substitution~(\ref{eq9}) leads to the invariant relations expressed
in the initial variables
\begin{equation}\label{eq16}
\begin{array}{l}
\displaystyle{(\xi _1^2  + \xi _2^2 )\omega _3  - 2[(\xi _1 \omega
_1 + \xi _2 \omega _2 )\alpha _3  + (\xi _2 \omega _1  - \xi _1
\omega _2 )\beta _3 ] = 0,} \\ [2mm]
 2[c(\xi _1^2  - \xi _2^2
) - (a^2  - b^2 )\xi _1 \xi _2 ](\omega _1^2  - \omega _2^2  + \xi
_1 ) +
\\[2mm]
\qquad+ [(a^2  - b^2 )(\xi _1^2  - \xi _2^2 ) + 4c\xi _1 \xi _2
](2\omega _1 \omega _2  + \xi _2 ) = 0.
\end{array}
\end{equation}
Here
$$
\xi _1  = \alpha _1  - \beta _2 , \quad \xi _2  = \alpha _2  + \beta
_1 .
$$
The equations of motion restricted to the manifold $N^4$ defined by
(\ref{eq16}) have two independent first integrals
\begin{equation}\label{eq17}
\begin{array}{l}
\displaystyle{H = \omega _1^2  + \omega _2^2  + \frac{1}{2}\omega
_3^2  - \alpha _1  - \beta _2  \equiv h,}
\\ [2mm]
K = (\omega _1^2  - \omega _2^2  + \alpha _1  - \beta _2 )^2  +
(2\omega _1 \omega _2  + \alpha _2  + \beta _1 )^2  \equiv k.
\end{array}
\end{equation}

The singular points of the initial system (\ref{eq2}), (\ref{eq3})
in this problem correspond to the body equilibria; there exists only
four such points. It follows from dynamics theorems that, if the
common level~(\ref{eq17}) does not contain any singular points and
the gradients of $H$ and $K$ are linearly independent, then each
connected component of~(\ref{eq17}) is a two-dimensional torus and
the trajectories on it satisfy the differential equations having the
last Jacobi multiplier. Therefore, by the time change, these
trajectories can be transformed to quasi-periodic ones.

As a result we have pointed out the two-parametric family (arbitrary
$h$ and $k$) of two-periodic motions of the rigid body in double
constant field under the conditions of Kowalevski type.

{\bf 3.\,The classical analogue.} Suppose that in the considered
problem ${\bs \beta} = 0$ (the second field vanishes). Then we
obtain the case of S.\,Kowalevski \cite{bib4}. The reduced phase
space of the Euler--Poisson variables has dimension 5. The invariant
relations (\ref{eq16}) take the form
\begin{eqnarray}
& \displaystyle{(\alpha _1^2  + \alpha _2^2 )\omega _3 - 2(\alpha _1
\omega _1  + \alpha _2 \omega _2 )\alpha _3  =
0,}\label{eq18} \\
& \displaystyle{2\alpha _1 \alpha _2 (\omega _1^2  - \omega _2^2  +
\alpha _1 ) - (\alpha _1^2  - \alpha _2^2 )(2\omega _1 \omega _2  +
\alpha _2 ) = 0}\label{eq19}
\end{eqnarray}

By means of the appropriate choice of the measurement unit $u_0$
make ${( {\bs \alpha},{\bs \alpha}) = 1}$. Write (\ref{eq19}) in the
form
$$
\displaystyle{ \frac{\omega _1^2 - \omega _2^2 + \alpha _1 }{\alpha
_1^2  - \alpha _2^2}=\frac{2\omega _1 \omega _2 + \alpha _2
}{2\alpha _1 \alpha _2}}
$$
and substitute to the Kowalevski integral to obtain
\begin{equation}\label{eq20}
\begin{array}{c}
\displaystyle{\omega _1^2  - \omega _2^2  + \alpha _1  = {\frac {\alpha _1^2  - \alpha _2^2 } {\alpha _1^2  + \alpha _2^2 }}\sqrt {k},}  \\
\displaystyle{2\omega _1 \omega _2  + \alpha _2  = {\frac{2\alpha _1
\alpha _2 } {\alpha _1^2  + \alpha _2^2 }}\sqrt {k}.}
\end{array}
\end{equation}
Recall the classical area integral
$$
L = \alpha _1 \omega _1  + \alpha _2 \omega _2  + \frac{1}{2}\alpha
_3 \omega _3  \equiv \ell.
$$
Calculate the combination $\Psi  = 2L^2  - H$,
\begin{equation}\label{eq21}
\begin{array}{l}
\displaystyle{\Psi  =  - \frac{1}{2} (\alpha _1^2  + \alpha _2^2
)[\omega _3 - 2{\frac{\alpha _1 \omega _1  + \alpha _2 \omega _2 }
{\alpha _1^2  + \alpha _2^2 }}]^2  +}   \\[2mm]
\displaystyle{\phantom{\Psi=} + \frac{1} {\alpha _1^2  + \alpha _2^2
} [(\alpha_1^2 - \alpha _2^2 )(\omega _1^2  - \omega _2^2  + \alpha
_1 ) + 2\alpha _1 \alpha _2 (2\omega _1 \omega _2  + \alpha _2 )].}
\end{array}
\end{equation}
Under the conditions (\ref{eq18}) and (\ref{eq20}) the last
expression turns into the following
\begin{equation}\label{eq22}
2\ell^2  - h = \sqrt {k}.
\end{equation}
Here the value $\sqrt k$ is algebraic. Eq.~(\ref{eq22}) defines the
2nd and 3rd classes of motions by the definition of
G.G.\,Appelrot~\cite{bib5, bib6}. Analyzing the structure of
(\ref{eq21}) and (\ref{eq22}), we conclude that the
three-dimensional manifold (\ref{eq18}), (\ref{eq19}) is exactly the
set of critical points of the combined first integral $(2L^2  - H)^2
- K$. In particular, one of the classic integrals on this manifold
becomes redundant. The other two define the closed orbits. These
are, naturally, the solutions on which one of the Kowalevski
variables remains constant. Emphasize that in the full phase space
including precession the corresponding motions are two-periodic for
almost all integral constants under the condition defined by
(\ref{eq22}).

Therefore the family of motions (\ref{eq16}) found in this paper
generalizes the so-called {\it especially remarkable} motions of the
2nd and 3rd Appelrot classes.

{\bf 4.\,Remarks.} After this paper was published in Russian journal
{\it Mekhanica tverdogo tela, 2002, N 32}, the author and
A.Y.\,Savushkin have received the separation of variables for the
class of motions found here \cite{bib7}. This separation provided
the explicit solutions in elliptic Jacobi functions and gave the
possibility to completely investigate the phase topology of the case
\cite{bib8}. In connection with the investigation of the set of
critical points and the arising bifurcation diagrams of the momentum
mapping of the Kowalevski top in double constant field, it was
proved by the author in \cite{bib9} that there exists only one more
case of Jacobi integrability, and for this new case the separation
of variables was found in \cite{bib10} leading to hyperelliptic
quadratures.

\end{document}